# Local Crystal Misorientation Influences Non-Radiative Recombination

*Sarthak Jariwala[1,2], Hongyu Sun[3], Gede W. P. Adhyaksa[3], Andries Lof[3], Loreta A. Muscarella[3], Bruno Ehrler[3], Erik C. Garnett[3], David S. Ginger[1]\**

[1] Department of Chemistry, University of Washington, Seattle, WA 98195, USA

[2]Department of Materials Science and Engineering, University of Washington, Seattle, WA 98195, USA

[3]Center for Nanophotonics, AMOLF, 1098 XG, Amsterdam, The Netherlands

*Corresponding Author: dginger@uw.edu

*Lead Contact: dginger@uw.edu

**Summary:**

We use ultrasensitive electron backscatter diffraction (EBSD) to map the local crystal orientations, grains, and grain boundaries in $CH_3NH_3PbI_3$ (MAPI) perovskite thin films. Although the true grain structure is broadly consistent with the morphology visible in scanning electron microscopy (SEM), the inverse pole figure maps taken with EBSD reveal grain structure and internal misorientation that is otherwise hidden. Local crystal misorientation is consistent with the presence of local strain which varies from one grain to the next. We acquire co-aligned confocal optical photoluminescence (PL) microscopy images on the same MAPI samples used for EBSD. We correlate optical and EBSD data, showing that PL is anticorrelated with the local grain orientation spread, suggesting that grains with higher degrees of crystalline orientational heterogeneity (local strain) exhibit more non-radiative recombination. We find that larger grains tend to have larger grain orientation spread, consistent with higher degrees of strain and non-radiative recombination.

**Introduction:**

Halide perovskite-based solar cells have experienced rapid gains in power conversion efficiency (PCE), with the current PCE record at 23.7% for single junction and 28% for Si-perovskite tandems.[1] Despite such high PCE for a material that is solution processed, the values are still well-below their realistically achievable PCE limits (~30% for single junction $CH_3NH_3PbI_3$[2] and ~32% for two terminal Si-perovskite tandems[3]). Non-radiative recombination, both in the film[4–7] and at the interfaces,[8,9] remains a major barrier to approach the radiative efficiency limits in halide perovskites. Our group has recently demonstrated that with surface passivation halide perovskites can achieve over 90% internal photoluminescence quantum efficiency (PLQE) and a quasi-Fermi level splitting approaching 97% of the radiative efficiency limit.[4,8] These results put halide perovskites alongside GaAs as one of the most radiatively efficient semiconductors.[10] However, intrinsic films (without surface passivation) have a significant source of non-radiative recombination (with internal PLQE ~ 10%).[4] A significant portion of this non-radiative recombination is thought to occur at under-passivated surfaces and grain boundaries.[4–6,11–13]

Unfortunately, the casual conflation of grain boundaries with morphological structures in halide perovskites has generated confusion about the nature of grain boundaries.[14,15] This situation has arisen because , currently, one of the most-used techniques to identify grains and grain boundaries in the halide perovskite community is scanning electron microscopy (SEM). However, while SEM provides information about the morphology of the film, conventional SEM does not provide any crystallographic information about the material. The crystallographic information is critical in the case of halide perovskites because domains as observed in the SEM may not accurately represent the grains and grain boundary terminations, leading to an overestimation of grain sizes.[16] Indeed, frustration with this topic has caused a leading solar cell researcher to tweet out "*SEM does not obvious[ly] [provide] information about 'grain size' or 'grain orientation' for metal halide perovskite active layers*".[17]

Traditionally, grain sizes, grain boundaries, and local crystal orientations in thin-film materials are mapped using electron backscatter diffraction (EBSD). However, this method has been notoriously difficult to apply to the halide perovskites most relevant for photovoltaics applications due to beam-induced damage.[16,18,19] Compared to SEM imaging, EBSD requires higher doses (higher current and/or longer integration time) to acquire sufficient contrast in the Kikuchi diffraction lines generated from backscattered electrons.[16] Therefore, the use of traditional EBSD detectors leads to significant beam-induced damage when characterizing halide perovskite solar cell materials.[16] Recently, Leonhard *et al.* reported that using low-vacuum mode with water pressures of 0.1-1 mbar in the sample chamber can minimize beam-induced damage and prevent surface charging due to ionization of the water molecules by the electrons during EBSD.[20] Some of us recently demonstrated the potential for a new solid-state EBSD detector to enable EBSD maps with high sensitivity, fast readout and without appreciable beam damage to the crystal structure.[21]

Understanding grain orientation and grain boundary properties has been critical to the development of many photovoltaic semiconductor technologies such as CdTe, GaAs, Cu(In$_{1-x}$Ga$_x$)Se$_2$, multicrystalline Si, and InP.[22–29] However, the role of grain orientation and grain boundary properties in halide perovskite semiconductors has remained poorly understood[30] despite the local optoelectronic heterogeneity observed at the microscale in these materials.[12,13,31,32] Here, we measure the local crystal orientation in solution-processed halide perovskite thin films using EBSD and demonstrate that the SEM morphology is not sufficient to characterize grain structure in these films. More importantly, we investigate the impact of grain-to-grain orientation heterogeneity and sub-grain orientation heterogeneity on local optoelectronic properties. We report the presence of local crystal misorientation leading to local strain within the grains and grain-to-grain orientation spread leading to strain heterogeneity within the film. Furthermore, we use crystallographic identification to unequivocally demonstrate and locate the sub-grain boundaries within individual grains. Lastly, we correlate EBSD and confocal photoluminescence measurements to measure the impact of local grain and sub-grain orientation heterogeneity on local photoluminescence. We observe higher non-radiative recombination in regions with higher orientation heterogeneity and lower non-radiative recombination in regions with lower orientation heterogeneity. Our results provide direct evidence for the impact of local crystal misorientation on the optoelectronic properties in halide perovskite thin films and point the way towards optimization of grain structure for improved photovoltaic performance.

**Results:**

Using a state of the art solid-state EBSD detector,[21] we optimized the beam current to be 100 pA at 6 kV accelerating voltage with a pixel integration time of 100 ms. We measure the Kikuchi diffraction patterns generated from the backscattered electrons using a traditional EBSD geometry with the electron beam hitting the sample at 70° with respect to the sample surface normal. Kikuchi diffraction lines generated at local Bragg angles contain the information about the local crystal orientation.[33,34] We scan across the sample and collect these Kikuchi patterns at every pixel and generate a local crystal orientation map of $CH_3NH_3PbI_3$ thin film.

Figure 1a shows an SEM image of a representative $CH_3NH_3PbI_3$ thin film grown by lead acetate and methylammonium based precursors with hypophosphorous acid additives (a recipe used for $CH_3NH_3PbI_3$-based photovoltaics, see Experimental Procedures for details)[35] and corresponding sharp Kikuchi patterns obtained at 6 kV, 10 kV and 20 kV. Consistent with X-ray diffraction (XRD) (Supplementary Figure S1), the sharp Kikuchi patterns indicate that these films are crystalline. Moreover, these results demonstrate that we can acquire sharp patterns at the relatively low accelerating voltage of 6 kV (acquired for only 100 ms per pixel) confirming the high sensitivity and fast readout of this new EBSD detector. Figure 1b and 1c show the SEM image of a $CH_3NH_3PbI_3$ film and the corresponding Inverse Pole Figure (IPF) map generated from fitting the Kikuchi patterns to a tetragonal $CH_3NH_3PbI_3$ structure[36,37] (see Supplementary Note S1 for details and Supplementary Figure S2 for comparison of indexed results to other phases) at every pixel, with an IPF color key. An IPF map shows the projection of the sample coordinate system (xyz) in the crystal coordinate system (abc). We note that the films have strong orientation along the [110] direction along the sample z-axis as evident from the Pole Density Figure (Supplementary Figure S3). Importantly, these data provide direct evidence for the crystallographic parameters and existence of grain boundaries in these $CH_3NH_3PbI_3$ thin films.

Broadly, we see that the SEM image in Figure 1b, and the EBSD image in Figure 1c are in relatively good agreement. Most of the "grains" observed in SEM are indeed also present in the EBSD image (Supplementary Figure S4 and Figure S5). However, while the features observed in SEM are generally consistent with grain structure obtained from EBSD measurements (Supplementary Figure S4 and Figure S5), *we emphasize that the EBSD maps provide critical information missing from the SEM images*. For example, based on SEM morphology alone, the black arrow in the SEM image in Figure 1b would be assigned as going across a single "grain boundary". However, from the corresponding IPF map (Figure 1c) we note that the film actually possesses 3 distinct boundaries separating 4 different local crystal orientation changes along the same arrow. Figure 1d depicts the changes in the local crystal orientation occurring across the arrow shown in Figure 1b and 1c (see Supplementary Figure S6 for higher resolution SEM image of Figure 1b and along the arrow). These results show that several large "grains" as seen in SEM image are composed of many smaller grains, similar to observations made using transmission electron microscope[38,39] and using EBSD in large grain (~50 μm) $CH_3NH_3PbBr_3$ samples.[21] Indeed, these differences between SEM and EBSD, perhaps provide another reason why efforts to relate morphological grain boundaries with carrier dynamics have not always found clear correlations.[15]

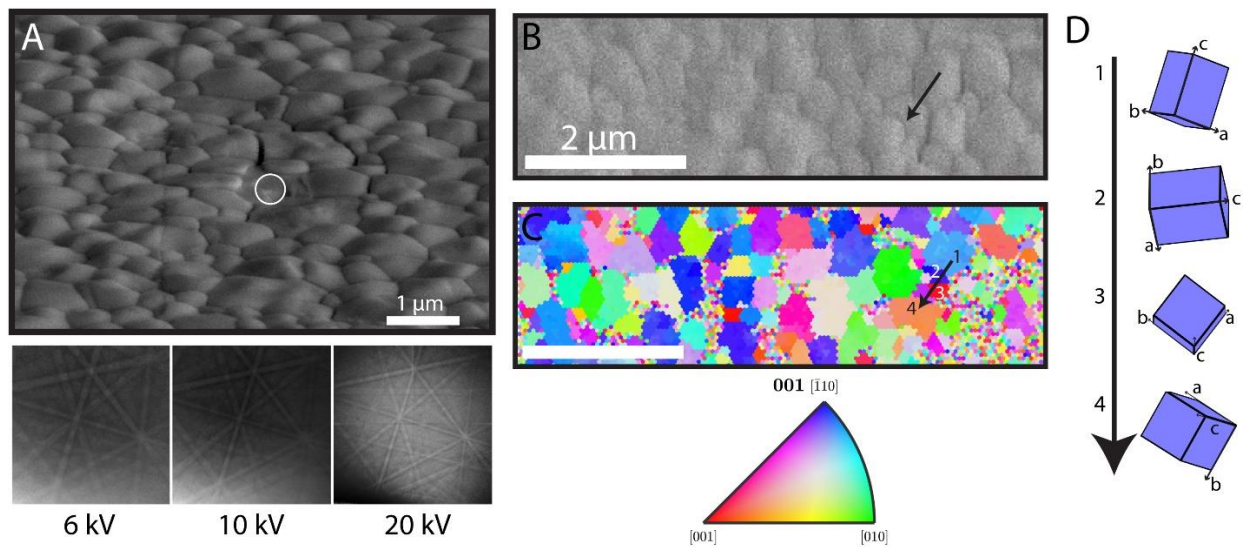

**Figure 1. Morphology and local crystal orientation measurements of $CH_3NH_3PbI_3$ thin films. (A) Scanning electron microscope (SEM) image of $CH_3NH_3PbI_3$ thin film tilted at 45° with representative sharp Kikuchi diffraction lines of a measurement point inside a grain (circled) collected in traditional electron backscatter diffraction (EBSD) geometry at 6 kV, 10 kV and 20 kV accelerating voltage. (B) SEM image and (C) Inverse Pole Figure (IPF) map generated from EBSD of a representative $CH_3NH_3PbI_3$ film with IPF color key. The 001 (in bold) in the IPF color key indicates that the IPF map plotted is along the sample z-direction. (D) Depiction of changes in local crystal orientation along the black arrow in (B) and (C) as viewed normal to the sample.**

**See Supplementary Figure S6 for a higher resolution SEM image of (B) and Figure S4 for other SEM image with IPF map.**

Figure 2a shows the inverse pole figure map of another typical $CH_3NH_3PbI_3$ thin film. By assigning an appropriate threshold for misorientation (4°, see Supplementary Figure S7 for results from different threshold values) we can assign each pixel in the inverse pole figure to a grain, and hence identify the grains and grain boundaries[40] (see Supplementary Note S1 and Figure S8 for more details on grain detection). Figure 2b shows the crystal grains identified by this approach, along with their grain boundaries. These maps provide a quantifiable identification of the grains and their orientation, revealing boundaries not visible in the SEM morphology images (see Supplementary Figure S9 for SEM image of the region in Figure 2).

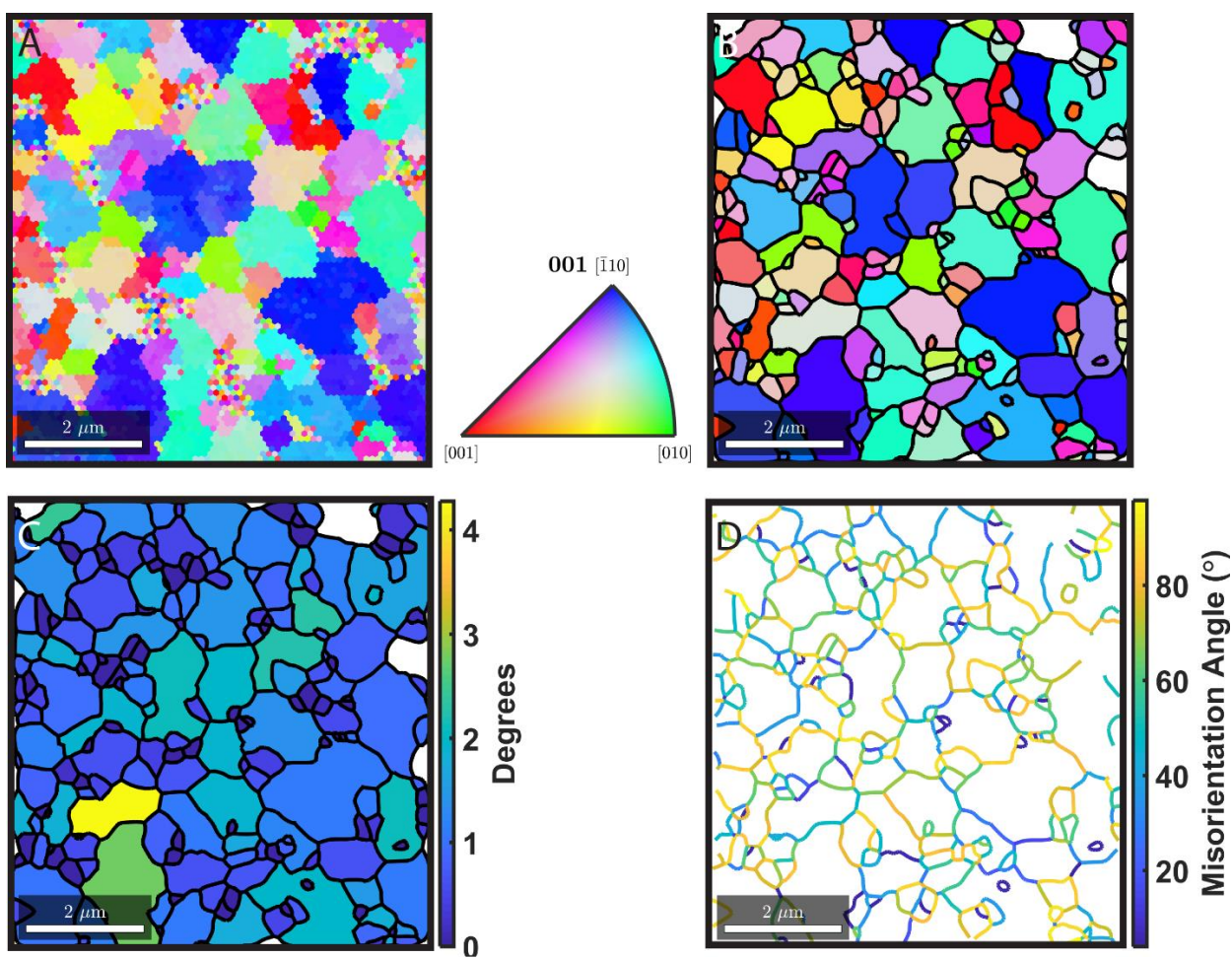

**Figure 2. Imaging the local orientation heterogeneity, grains and grain boundaries in $CH_3NH_3PbI_3$ thin films using electron back scatter diffraction (EBSD). (A)** Inverse Pole Figure (IPF) map of $CH_3NH_3PbI_3$ generated using EBSD and IPF color key. **(B)** Grains identified from IPF map with a 4° orientation threshold and plotted with their mean orientation. **(C)** Plot of grain orientation spread (GOS) showing grain-to-grain heterogeneity in average local misorientation in the same film. **(D)** Grain boundary network with their respective misorientation angles showing the degree of misorientation between two neighboring grains. (B), (C) and (D) have the same orientation threshold (4°). See Supplementary Figure S9 for SEM image of the region.

In addition to identifying grains and grain boundaries, the analysis of the IPF data in Figure 2 provides an additional level of insight regarding crystallographic misorientation within individual grains. For example, we observe local variations in crystal orientations within individual grains (Supplementary Figure S10) across the entire film that exceed the orientation noise in local crystallographic orientation obtained from fitting the Kikuchi pattern (see Supplementary Note S1 for details). Figure 2b plots the grains identified in Figure 2a, color coded according to their mean orientation.[41,42] The average deviation in orientation between each pixel (point) within a grain and the mean grain orientation of that grain is defined as the grain orientation spread (GOS).[43] Figure 2c plots the GOS values in the same region as Figure 2a and 2b of this representative $CH_3NH_3PbI_3$ thin film. Importantly, we observe a range of grain-to-grain orientational heterogeneity in the film with GOS values as low as 0° (perfectly ordered) and as high as 4.3° of orientation spread within individual grains. These grain-to-grain local variations in crystal orientation are indicative of local strain distributions within the grains.[43] Higher GOS values indicate grains with higher strain, and lower GOS values indicate grains with lower strain.[43] We note that the misorientation within the grain and between neighboring points observed is not a consequence of beam induced degradation in the material (see Supplementary Figure S11). These results demonstrate the existence of potentially important local structural and strain heterogeneity in halide perovskite thin films stemming from the local crystal orientation variations. These results are consistent with, but provide better resolution, than those of Stranks *et al.,* who recently reported strain heterogeneity at long length scales in halide perovskite thin films based on μ-XRD measurements.[38] Importantly, our results further demonstrate that the local strain heterogeneity observed in halide perovskite semiconductors originate from local crystal misorientation within grains. In most semiconductors, local lattice imperfections, such as local crystal orientation changes within grains observed here, can act as non-radiative recombination centers,[44] which in turn would have a significant influence on the performance of perovskite solar cells. We will explore this possibility in more detail below.

Alongside local crystal orientation heterogeneity, orientation imaging using EBSD also allows us to probe and understand the nature of grain boundaries in halide perovskites. Misorientation between two neighboring grains (grain boundary misorientation) can be quantified using the crystallographic misorientation angles and misorientation axes obtained using EBSD. The misorientation angle is the angle by which a grain is rotated, about the misorientation axis, such that the orientation of the transformed grain matches the neighboring grain after transformation. Figure 2d shows the grain boundary network and their respective misorientation angles between neighboring grains. The misorientation angle/axis statistics (Supplementary Figure S12 and Figure S13) demonstrate the presence of high angle grain boundaries with high frequency and strong preferred orientation of misorientation axis along the [110] direction. Importantly, Figure 2d shows that different grain boundaries have different misorientation and therefore, may have different properties. In other words, these films exhibit a range of different grain boundaries, which might be expected to exhibit different properties. As stated earlier, understanding grain boundary properties has been critical to the development of many other photovoltaic semiconductor technologies such as CdTe, GaAs, InP, and Cu(In$_{1-x}$Ga$_x$)Se$_2$,[22–29] and indeed other materials, such as high temperature polycrystalline superconductors.[45,46] We anticipate that understanding the nature of the different grain boundaries identified by EBSD will be critical to understanding the properties of halide perovskite photovoltaics,[21] and we speculate that properties such as anisotropic charge carrier transport across grain boundaries observed in halide perovskites[13,47–49] could depend on grain boundary properties such as misorientation angle, type of boundary, and the boundary interface energy.

Next, to investigate the structural heterogeneity at a sub-grain level in more detail, we examine crystal orientations within an individual grain. Due to high coefficient of thermal expansion anisotropy in tetragonal CH$_3$NH$_3$PbI$_3$, phase transition from cubic to tetragonal CH$_3$NH$_3$PbI$_3$ during film formation can introduce a significant amount of local residual stresses in the material.[50,51] These micro-stresses may be concentrated at the grain boundaries and sub-grain boundaries within the grain. Figure 3a shows the inverse pole figure of an individual grain from a CH$_3$NH$_3$PbI$_3$ film with sub-grain boundaries. The grains are identified using 4° as the threshold, however, the grain represented in Figure 3 is not affected by different thresholds (see Figure S14). Sub-grain boundaries have been shown to influence charge carrier recombination in other semiconductor photovoltaic technologies like GaAs and InP.[27,28] Recently, using optical microscopy Li *et al.* suggested the presence of sub-grain special boundaries in large grain NH$_2$CHNH$_2$PbI$_3$ thin films.[49] Here, we use crystallographic identification to unequivocally demonstrate the existence of sub-grain boundaries and precisely determine their location within the grain.

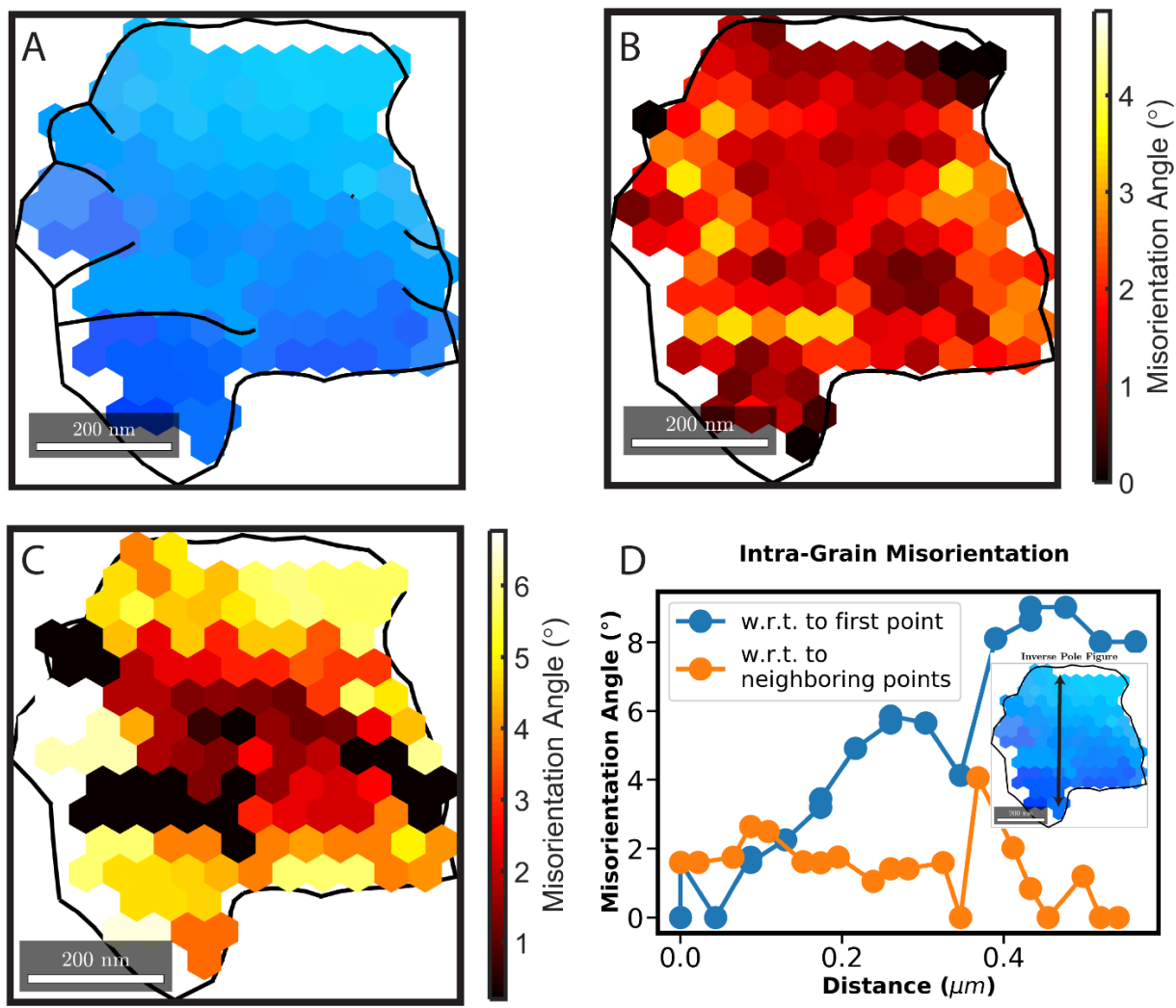

**Figure 3. Sub-grain boundaries and intra-grain misorientation. (A)** Inverse Pole Figure (IPF) of an individual grain in a $CH_3NH_3PbI_3$ thin film with sub-grain boundaries. The outer black lines represent the outer grain boundary (i.e. with other grains) and the inner black lines represent the sub-grain boundaries. The IPF color key is the same as Figure 1 and 2. The grains are identified using 4° as the threshold. **(B)** Kernel average misorientation (KAM) showing the degree of misorientation between neighboring pixels within the grain. **(C)** Misorientation with respect to mean orientation of the grain showing the local orientation heterogeneity in the grain. **(D)** Plot of misorientation angle as a function of distance along the black arrow in the IPF in the inset. Blue circles denote the misorientation of each point with respect to the first point along the black arrow and orange circles denote the misorientation of each point with respect to its neighboring points along the arrow.

As stated earlier, local variations in lattice orientation can manifest as residual strain within the material.[50] Here, we analyze local variations in lattice orientations to infer the local strain within an individual grain. Kernel average misorientation (KAM) provides the average misorientation between a measurement point and its neighboring points.[43] In our case, it provides the average misorientation between each point and its six neighboring points within the same grain. We note that the KAM is a kernel or an individual measurement property where each point in the grain has a local misorientation value whereas the previously described GOS is a grain property where the entire grain is assigned a local misorientation value. Figure 3b shows the KAM value calculated and plotted for each point within the grain. We observe KAM values as low as 0° and as high as 4.87° corresponding to low and high local misorientation regions respectively. We note that the range of KAM values (0 to 4.87°) observed here are *within* the same grain, demonstrating significant strain heterogeneity within an individual grain. Interestingly, the sub-grain boundaries (Figure 3a) are near regions with high KAM values (Figure 3b) consistent with the model that high misorientations within a grain will lead to sub-grain boundaries. Our observations show that local crystal misorientation generates the local strain heterogeneity observed within individual grains.

As evident from Figure 3a and 3b, the selected grain has a distribution of orientations within itself, exhibiting a GOS of 3.41°. Figure 3c shows the misorientation of individual measurement points within a grain with respect to the mean orientation of the grain. Regions near to the reference mean orientation have lower misorientation angle and vice-versa. We observe higher misorientation angles with respect to mean orientation closer to the grain boundaries suggesting a higher degree of lattice rotation or lattice bending near grain boundaries.

Figure 3d shows the plot of intra-grain misorientation angle as a function of distance within the grain, along the black arrow in the inset figure. Importantly, Figure 3d shows that the misorientation from one end of the grain to the other end can be as high as 9° (even though the grain boundaries were identified using a threshold of 4°). This result could indicate the possibility of variations in carrier transport both across, and within, an individual grain if the strain affects the local carrier mobility. We note that the sub-grain heterogeneity present in the grain shown in Figure 3 is representative of the population (See Supplementary Figure S15 and Figure S16 for SEM image of an individual grain along with sub-grain heterogeneity characterization and sub-grain heterogeneity in more grains from different films, respectively).

Finally, having quantified the local grain orientation spread, and hence the local strain, we use our high-resolution EBSD data set to explore whether grain-to-grain variations in near-surface strain (Supplementary Figure S17) could be sources of non-radiative recombination within the perovskite films. We expect the defects (and the strain) to be dominated by the surface, because surface passivated halide perovskite thin films have demonstrated single crystal like quality.[4–6,52] To this end, we overlay the grain-boundary network obtained from grain identification onto the confocal photoluminescence (PL) image obtained on the same region (Figure 4a). The grain boundary network and the confocal PL image were aligned using an image registration program and fiducial marks created using Atomic Force Microscope tip (see Supplementary Note S2 for details).

To look for any relationship between local orientation heterogeneity and local non-radiative recombination, we plot photoluminescence intensity as a function of grain orientation spread. Our group has previously demonstrated that confocal PL maps, when measured at excitation densities below the local trap density (~$10^{15}$ cm$^{-3}$ to ~$10^{17}$ cm$^{-3}$),[12,53–55] can reflect the local trap distribution in the film.[12,13] Although carrier diffusion is still important,[13,56] under such conditions, high PL intensities correspond to regions with lower trap density, and low PL intensities to regions with higher trap density.[13]

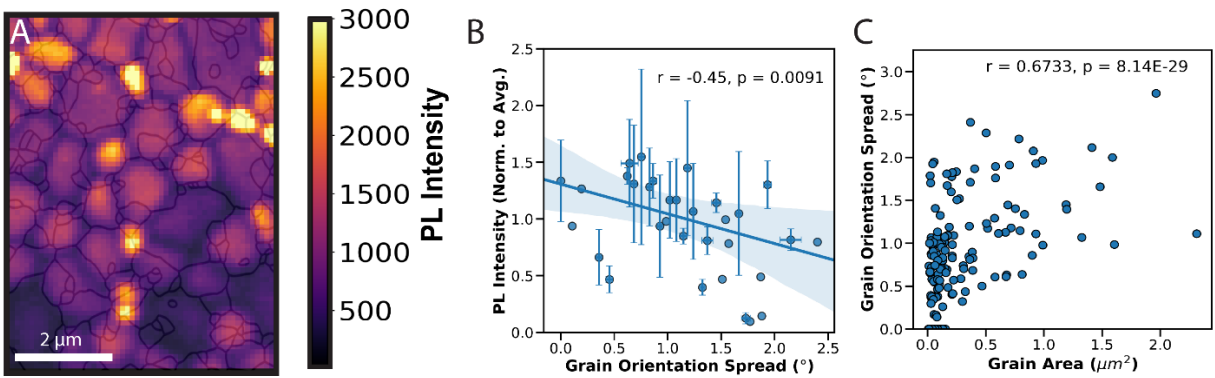

**Figure 4. Correlation between local luminescence, orientation heterogeneity and grain area in CH$_3$NH$_3$PbI$_3$. (A) Aligned confocal photoluminescence (PL) image and grain boundary network of the same region. (B) Plot of PL intensity as a function of Grain Orientation Spread (GOS) in MAPI showing negative correlation, with high statistical significance (p=0.0091). The grains are binned in GOS intervals. The line represents a linear regression fit to the data with the shaded region representing a 95% confidence interval for the regression. Error bars represent the standard deviation of average PL intensity in a specific GOS interval. (C) Plot of grain orientation spread as a function of grain area showing positive correlation, with high statistical significance (p=8.14e-29). See Supplementary Note S3 for discussion on p-value and statistical significance.**

Here, we measure confocal PL maps at low incident excitation fluences of ~0.2 µJ/cm$^2$ per pulse (at 470 nm with 1 MHz repetition rate) corresponding to an initial carrier excitation density of ~$10^{16}$ cm$^{-3}$ and time-averaged carrier density of ~$10^{15}$ cm$^{-3}$. Next, we measure the local crystal orientation across the same region using EBSD as demonstrated earlier. Using this approach, Figure 4b plots the normalized average PL intensity within a grain versus the grain orientation spread within the corresponding grain for more than 100 grains from different samples (see Supplementary Figure S18 for correlation plots from individual sample sets). We find a negative correlation between the grain orientation spread and PL intensity with high statistical significance (p-value = 0.0091, Supplementary Note S3 for more details on p-value). In other words, the larger the grain orientation spread, the lower the PL intensity of the local perovskite. This result provides a direct link between local intra-grain strain and local nonradiative defect density by showing that grains with higher grain orientation spreads have lower PL intensity and therefore, higher trap densities contributing to non-radiative recombination.

Figure 4c shows the relation between grain area and the corresponding orientation spread within the grain. We find a strong positive correlation between the grain area and grain orientation spread, with high statistical significance (p-value = 8.14e-29). We note that this conclusion does not depend on the threshold value chosen during grain detection. In other words, we observe strong positive correlation between grain area and grain orientation spread for all different grain threshold values (Supplementary Figure S19). This means that, in the case of the samples measured in this study, as the grain size increases the orientation spread (and hence, local strain) also increases. We note that at fluences used in this study (corresponding to trap densities in the film) we find very low anticorrelation (r = -0.2) between PL intensity and grain area (Supplementary Figure S20), consistent with our previous observations.[13] Importantly, this result indicates that the correlation between PL and grain orientation spread shown in Figure 4b is *not* solely due to the relation between grain size and PL, instead grain orientation spread appears to be more important than grain size.

**Discussion:**

Notably, an increase in grain orientation spread (and local strain) with increasing grain area might help explain why some of the smaller grain films in the literature have shown higher PCE, and why increasing *apparent* grain sizes, as imaged by SEM, has not always corroborated with increasing PCE.[57] Specifically, we speculate that smaller grain films studied to date may have less orientation spread and thus, lower strain compared to large grains.

However, we emphasize that these results do not indicate researchers should necessarily aim to create perovskite films with smaller grains for solar cell applications. Grain boundaries and surface traps act as centers for non-radiative recombination in many materials, including halide perovskites.[5–7,12,13,21,53,58–65] Instead, the point we are making is that local crystal orientation is also important, perhaps more so than grain size in some films. Ideally, we would want films with larger grain sizes and low orientation spread within a grain, i.e. highly oriented and low strain. Promisingly, Choi and co-workers recently demonstrated that perovskite thin films with higher degrees of orientation can be achieved at the lab scale.[66,67] As lab-scale deposition techniques are translated to large-scale deposition methods such as roll-to-roll printing, spray coating and slot die coating;[68,69] our results, and the general methods herein, provide insight for optimizing deposition conditions, and suggest a need for orientation and strain engineering.[70]

In conclusion, we demonstrate local crystal orientation maps of archetypal $CH_3NH_3PbI_3$ perovskite thin films, most commonly investigated for solar cell applications, generated using the state of the art EBSD detector with high sensitivity and fast readout. We show that while SEM morphology is generally consistent with EBSD, there are features in the film that are only visible in EBSD and therefore, SEM morphology alone is insufficient to identify grains and grain boundaries. In addition to providing true grain sizes for halide perovskites, orientation imaging using EBSD also allows us to probe the crystallographic nature of the film. We observe local crystal orientation heterogeneity within individual grains throughout the thin film and we find that the orientation heterogeneity varies from grain-to-grain. Our results demonstrate that although a grain is defined as a unit of microstructure with a single orientation, in reality, in perovskites a grain has an average orientation with significant orientational spread or heterogeneity which contributes to the local strain within the grain. We propose that such orientational heterogeneity within grains could be a result of heterogeneous nucleation and growth, variations in local concentration gradients[71] and/or phase transitions post-annealing.[37,50,51,72,73] Next, by studying the crystallographic nature of grain boundaries, we show that there are a range of different grain boundary misorientations and therefore, different grain boundaries will have different properties such as carrier transport, interface energy, etc. Using EBSD, we also demonstrate the presence of sub-grain boundaries formed near regions of high local crystal misorientation. We observed misorientations as high as 9° from one end of the grain to another. Finally, we use correlated confocal PL microscopy and EBSD to understand the relationship between local crystal orientation heterogeneity and local non-radiative recombination, as it relates directly to photovoltaic performance. We observe low PL intensity (high non-radiative recombination) in regions with high grain orientation spreads (high strain regions) and high PL intensity (low non-radiative recombination) in regions with low grain orientation spreads (low strain regions). Furthermore, we observe that the orientation spread increases with increasing grain sizes. Our results point towards the need to understand and control the local crystal orientation by engineering deposition protocols that provide large grains with low orientation heterogeneity (lower strain). These results provide critical insight into the interplay between local crystal orientation heterogeneity and local non-radiative recombination and furthers our understanding of the local structure-function interplay in halide perovskite thin films for photovoltaics.

**Experimental Procedures:**

**Precursor Preparation and Film Deposition**

In a nitrogen-filled glovebox, a methylammonium iodide (MAI) solution (1.78 M) was made by dissolving MAI (Dyesol, CAS:14965-49-2) in anhydrous N,N-dimethylformamide (DMF). Lead acetate trihydrate (PbOAc$_2$ ·3H$_2$O) (99.999%, Sigma-Aldrich, CAS:6080-56-4) was added to the MAI solution at a 3:1 molar ratio of MAI to PbOAc$_2$ ·3H$_2$O (0.59 M). A hypophosphorous acid (HPA) solution was further added to the precursor solution with a molar ratio HPA/PbOAc2·3H2O of 8%.[35]

Glass substrates were cleaned by sequentially sonicating in 2% Micro-90 detergent, DI water, acetone, then propan-2-ol. Prior to film deposition, the glass substrates were plasma-cleaned. The perovskite precursor solution was spin-coated on top in a nitrogen-filled glovebox, at 2000 rpm for 45 s to form the perovskite thin film layer. The films were then annealed at room temperature for 10 mins and at 100°C for 5 mins.

**Fluorescence Lifetime Imaging Microscopy**

A custom scanning confocal microscope built around Nikon TE-2000 inverted microscope with a sample stage controlled by Physik Instrumente E-710 piezo controller was used to measure fluorescence images. The system was first calibrated using 200 nm fluorescent microspheres (Lifetechnologies FluoSpheres Polystyrene Microspheres, 200 nm, red fluorescent, 580/605 nm). The sample was illuminated, through a LU Nikon Plan Fluor 100x infinity corrected dry objective (0.9 NA), with 470 nm pulsed diode laser (PDL-800 LDH-P-C-470B, 1 MHz, ~300 ps pulse width) and the emission was filtered using a 50/50 dichroic beam splitter and a pair of 500/600 nm long-pass filters. The local fluorescence was measured by directing the emission to a Micro Photon Devices PDM series single-photon avalanche photodiode with a 50 µm active area. The pixel size (or scanning step size) used for fluorescence lifetime images was 100 nm and the pixel dwell time (or integration time) was 100 ms.

**Electron Back-Scatter Diffraction (EBSD)**

PELCO® conductive silver paint was first deposited on top of the perovskite film on glass to cover all the edges of the film. Upon drying, it forms a thick silver layer around edges to provide good conductivity with a sheet resistance of 0.066 ohm/sq. The dried silver paint was then connected to a metal sample holder for further grounding. EBSD was performed in the region between the silver paint (~5 mm x 5mm) to avoid any charging effects. The electron beam was optimized to 100 pA beam current and 6 kV accelerating voltage with a 10 mm working distance to the sample. The sample was tilted such that the electron beam was hitting the sample at 70° with respect to the sample surface normal. The electron beam was scanned across the sample with a step size of 50-100 nm to collect Kikuchi patterns at every pixel. The patterns were collected using EDAX OIM software and the pixel integration time was 100 ms. The electron dose rate for the measurement was estimated to be ~0.1 electrons/Å$^2$ per second. The total dose for 8µm X 8µm scan with 0.1µm step size and 100 ms pixel dwell time was estimated to be ~62 electrons /Å$^2$. The 1/e decay length for the backscatter electron energy distribution in the X-Y plane is ~106 nm (Figure S17), thus justifying the assumption in the calculation that the beam size is similar to the step size in the measurement.

**Acknowledgements:**

This paper is primarily based on research supported by the DOE DE-SC0013957. S.J. acknowledges support from the University of Washington Clean Energy Institute and the National Science Foundation Research Traineeship under Award NSF DGE-1633216. D.S.G acknowledges support from the University of Washington, Department of Chemistry Kwiram Endowment. G.W.P.A. was supported by TKI Urban Energy, "COMPASS" (TEID215022). E.C.G. and G.W.P.A. received funding from the European Research Council grant No. 337328 "NanoEnabledPV". E.C.G. and H.S. were supported by the Dutch Science Foundation (NWO) through the Joint Solar Program III. L.A.M. and B.E. were supported by the Dutch Science Foundation (NWO) through the project No. 680-47-553.

**Author Contributions:**

S.J. fabricated the films for confocal PL imaging and EBSD, carried out the PL experiments and analyzed the EBSD and PL results and coordinated the experiments under the supervision D.S.G. H.S., G.W.P.A., A.L. and L.A.M. carried out the EBSD measurements and indexing under the supervision of E.C.G. and B.E. D.S.G. and S.J. conceived the idea, designed the experiments, interpreted the results and wrote the manuscript. All authors contributed to the editing of the manuscript.

**Declaration of Interests:**

The EBSD detector used in this study is now being commercialized by Amsterdam Scientific Instruments.

[62] Brenes, R., Eames, C., Bulović, V., Islam, M.S., and Stranks, S.D. (2018). The Impact of Atmosphere on the Local Luminescence Properties of Metal Halide Perovskite Grains. Adv. Mater. *30*, 1–8.

[63] Bischak, C.G., Sanehira, E.M., Precht, J.T., Luther, J.M., and Ginsberg, N.S. (2015). Heterogeneous Charge Carrier Dynamics in Organic-Inorganic Hybrid Materials: Nanoscale Lateral and Depth-Dependent Variation of Recombination Rates in Methylammonium Lead Halide Perovskite Thin Films. Nano Lett. *15*, 4799–4807.

[64] Draguta, S., Thakur, S., Morozov, Y. V, Wang, Y., Manser, J.S., Kamat, P. V., and Kuno, M. (2016). Spatially Non-uniform Trap State Densities in Solution-Processed Hybrid Perovskite Thin Films. J. Phys. Chem. Lett. *7*, 715–721.

[65] Vrucinic, M., Matthiesen, C., Sadhanala, A., Divitini, G., Cacovich, S., Dutton, S.E., Ducati, C., Atature, M., Snaith, H., Friend, R.H., et al. (2015). Local Versus Long-Range Diffusion Effects of Photoexcited States on Radiative Recombination in Organic-Inorganic Lead Halide Perovskites. Adv. Sci. *2*, 1–6.

[66] Foley, B.J., Cuthriell, S., Yazdi, S., Chen, A.Z., Guthrie, S.M., Deng, X., Giri, G., Lee, S.-H., Xiao, K., Doughty, B., et al. (2018). Impact of Crystallographic Orientation Disorders on Electronic Heterogeneities in Metal Halide Perovskite Thin Films. Nano Lett. *18*, 6271–6278.

[67] Chen, A.Z., Foley, B.J., Ma, J.H., Alpert, M.R., Niezgoda, J.S., and Choi, J.J. (2017). Crystallographic orientation propagation in metal halide perovskite thin films. J. Mater. Chem. A *5*, 7796–7800.

[68] Dou, B., Whitaker, J.B., Bruening, K., Moore, D.T., Wheeler, L.M., Ryter, J., Breslin, N.J., Berry, J.J., Garner, S.M., Barnes, F.S., et al. (2018). Roll-to-Roll Printing of Perovskite Solar Cells. ACS Energy Lett. *3*, 2558–2565.

[69] Yang, M., Li, Z., Reese, M.O., Reid, O.G., Kim, D.H., Siol, S., Klein, T.R., Yan, Y., Berry, J.J., Van Hest, M.F.A.M., et al. (2017). Perovskite ink with wide processing window for scalable high-efficiency solar cells. Nat. Energy *2*, 17038.

[70] Zhu, C., Niu, X., Fu, Y., Li, N., Hu, C., Chen, Y., He, X., Na, G., Liu, P., Zai, H., et al. (2019). Strain engineering in perovskite solar cells and its impacts on carrier dynamics. Nat. Commun. *10*, 815.

[71] Hu, Q., Zhao, L., Wu, J., Gao, K., Luo, D., Jiang, Y., Zhang, Z., Zhu, C., Schaible, E., Hexemer, A., et al. (2017). In situ dynamic observations of perovskite crystallisation and microstructure evolution intermediated from [PbI 6 ]4- cage nanoparticles. Nat. Commun. *8*.

[72] Vorpahl, S.M., Giridharagopal, R., Eperon, G.E., Hermes, I.M., Weber, S.A.L., and Ginger, D.S. (2018). Orientation of Ferroelectric Domains and Disappearance upon Heating Methylammonium Lead Triiodide Perovskite from Tetragonal to Cubic Phase. ACS Appl. Energy Mater. *1*, 1534–1539.